\documentclass[times]{aastex631}
\usepackage{amssymb}
\usepackage{amsmath}
\usepackage{gensymb}
\usepackage{graphicx}
\usepackage{xcolor}
\usepackage{natbib}
\usepackage{float}

\newcommand{\beq}{\begin{equation}}
\newcommand{\eeq}{\end{equation}}

\newcommand{\Ms}{\textrm{M}_*}
\newcommand{\Msun}{\textrm{M}_\odot}
\newcommand{\kmps}{km~s$^{-1}$}
\newcommand{\MHI}{\rm{M_{H{\textsc i}}}}

\newcommand{\htwo}{{\rm H_2}}

\newcommand{\hi}{H{\sc i}}

\newcommand{\hii}{H{\sc i} 21\,cm}
\defcitealias{NatPaper}{vD18}
\graphicspath{{./}{figures/}}

\shorttitle{H{\sc i} 21\,cm Emission at $z \approx 1.0$}
\shortauthors{Chowdhury, Kanekar and Chengalur}

\begin{document}
	
	\title{Insufficient Gas Accretion Caused the Decline in Cosmic Star-Formation Activity 8 Billion Years Ago}

	\correspondingauthor{Aditya Chowdhury}
	\email{chowdhury@ncra.tifr.res.in}
	
	\author{Aditya Chowdhury}
	\affil{National Centre for Radio Astrophysics, Tata Institute of Fundamental Research, Pune, India.}
	
	\author{Nissim Kanekar}
	\affil{National Centre for Radio Astrophysics, Tata Institute of Fundamental Research, Pune, India.}
	
	\author{Jayaram N. Chengalur}
	\affil{National Centre for Radio Astrophysics, Tata Institute of Fundamental Research, Pune, India.}

	
	
	\begin{abstract}
	Measurements of the atomic hydrogen (H{\sc i}) properties of high-redshift galaxies are critical to understanding the decline in the star-formation rate (SFR) density of the Universe after its peak $\approx8-11$ Gyr ago. Here, we use $\approx510$~hours of observations with the upgraded Giant Metrewave Radio Telescope to measure the dependence of the average H{\sc i} mass of star-forming galaxies at $z=0.74-1.45$ on their average stellar mass and redshift, by stacking their H{\sc i} 21~cm emission signals. We divide our sample of 11,419 main-sequence galaxies at $z= 0.74-1.45$ into two stellar-mass ($\textrm{M}_*$) subsamples, with $\textrm{M}_*>10^{10}~\textrm{M}_\odot$ and $\textrm{M}_*<10^{10}~\textrm{M}_\odot$, and  obtain clear detections, at $>4.6\sigma$ significance, of the stacked H{\sc i} 21~cm emission in both subsamples. We find that galaxies with $\textrm{M}_*>10^{10}~\textrm{M}_\odot$,  which  dominate the decline in the cosmic SFR density at $z\lesssim1$, have H{\sc i} reservoirs that can sustain their SFRs for only a short period, $0.86\pm0.20$~Gyr, unless their H{\sc i} is replenished via accretion. We also stack the H{\sc i} 21~cm emission from galaxies in two redshift subsamples, at $z=0.74-1.25$ and $z=1.25-1.45$, again obtaining clear detections of the stacked H{\sc i} 21~cm emission signals, at $>5.2\sigma$ significance in both subsamples. We find that the average H{\sc i} mass of galaxies with $\langle\textrm{M}_* \rangle\approx10^{10}~\textrm{M}_\odot$ declines steeply over a period of $\approx1$~billion years, from $(33.6\pm6.4) \times 10^9~\textrm{M}_\odot$ at $\langle z\rangle\approx1.3$ to $(10.6\pm1.9)\times10^9~\textrm{M}_\odot$ at $\langle z\rangle\approx1.0$, i.e. by a factor $\gtrsim3$. We thus find direct evidence that accretion of H{\sc i}\ onto  star-forming galaxies at $z\approx1$ is insufficient to replenish their H{\sc i} reservoirs and sustain their SFRs, thus resulting in the decline in the cosmic SFR density 8 billion years ago.

	\end{abstract}
	
	\keywords{Galaxy evolution --- Radio spectroscopy --- Neutral hydrogen clouds}
	
	\section{Introduction}
  Understanding the evolution of the star-formation activity in galaxies is a key open issue in galaxy evolution \citep[e.g.][]{Madau14}. The cosmic comoving star-formation rate (SFR) density of the Universe is known to peak in the redshift range $z\approx 1-3$, and to then decline by an order of magnitude over the next $\approx 8$ billion years, to its value today \citep[e.g.][]{LeFloch05,Bouwens10}. Approximately $ 90\%$ of cosmic star-formation at $z\lesssim 2.5$ takes place in galaxies lying on the galaxy ``main sequence'', a power-law relation between the SFR and the stellar mass followed by most star-forming galaxies at any redshift \citep[e.g.][]{Noeske07,Rodighiero11}. The ``amplitude'' of the main sequence, i.e. the SFR at a fixed stellar mass, has been found to decline by a factor of $\approx 10$ between $z\approx1$ and $z\approx0$ \citep[e.g.][]{Whitaker14}, similar to the decline in the cosmic SFR density over the same period. The cause of the decline in the star-formation activity of the Universe from $z\approx1$ to $z\approx0$ remains unknown today. Addressing this issue requires us to understand the evolution of the gas mass of galaxies, the fuel from which the stars form. 
  
  The primary fuel for star-formation is neutral atomic hydrogen (\hi). The atomic gas content of galaxies is best measured with the hyperfine \hii\ transition in the \hi\ ground state, at a wavelength of $\approx 21.11$~cm.  Unfortunately, the low Einstein A-coefficient of this transition has meant that even deep observations with the best radio telescopes today are only able to detect \hii\ emission from individual galaxies at low redshifts, $z \lesssim 0.4$ \citep[e.g.][]{Fernandez16}. However, the {\it average} \hi\ mass of a sample of galaxies can be measured by ``stacking'' the \hii\ emission signals from a large number of galaxies with known spatial positions and redshifts \citep[e.g.][]{Zwaan00,Chengalur01,Lah07,Kanekar16,Bera19}. Such \hii-stacking studies have been used to estimate the average \hi\ mass of star-forming galaxies out to $z \approx 1.3$ \citep{Chowdhury20,Chowdhury21}.
  
  Recently, \citet{Chowdhury20} applied the \hii-stacking technique to galaxies in the DEEP2 fields to obtain the first detection of the average \hii\ emission signal from star-forming galaxies at $z\gtrsim1$ \citep[see also][]{Chowdhury21}. Their results suggested that a decline in the \hi\ reservoir of star-forming galaxies at $z \gtrsim 1$ could account for the observed decline in the SFR density of the Universe at later times. 
  
  In this \emph{Letter}, we report first results from the Giant Metrewave Radio Telescope (GMRT) Cold-\hi\ AT $z\approx1$ (CAT$z$1) Survey, a 510-hr upgraded GMRT \hii\ emission survey of the DEEP2 fields, designed to directly probe the cause of the decline in the cosmic SFR density, by measuring redshift evolution in the average \hi\ mass of star-forming galaxies around $z\approx1$ when the star-formation activity of the Universe begins to decline rapidly.\footnote{Throughout this work, we use a flat ``737'' Lambda-cold dark matter cosmology, with $\Omega_m=0.3$, $\Omega_\Lambda = 0.7$, and $H_0 = 70$~\kmps~Mpc$^{-1}$.}
  

	\section{Observations and Data Analysis}
	 The GMRT-CAT$z$1 survey is a deep \hii\ emission survey of galaxies at $z=0.74-1.45$ with the upgraded GMRT \citep{Swarup91,Gupta17} Band-4 receivers, targetting three fields of the DEEP2 Galaxy Redshift Survey \citep{Newman13}. 
	 The survey uses the \hii\ stacking approach to carry out the first characterisation of the \hi\ properties of galaxies at $z\approx1$. The DEEP2 Survey is uniquely suited to this experiment because it provides high-accuracy \citep[$\approx 55$~\kmps; ][]{Newman13} measurements of the spectroscopic redshifts of galaxies at $z=0.74-1.45$, whose \hii\ lines are redshifted into the GMRT Band-4 frequency coverage ($\approx 550-830$~MHz), and which are located in sky regions that are well-matched to the the GMRT Band-4 field of view \citep[e.g.][]{Kanekar16,Chowdhury20,Chowdhury21}. The survey design, the observations, and the analysis of the GMRT-CAT$z$1 survey data are described in detail in \citet{Chowdhury22}. We provide here a brief summary of the observations and the analysis procedure.
	
	The survey was carried out over three GMRT observing cycles, using 90~hours in October 2018 -- March 2019 \citep[presented in][]{Chowdhury20}, 170~hours in  October 2019 -- March 2020, and  250~hours in May 2020 -- October 2020 (proposals 35\_087, 37\_063, and 38\_033; PI: A. Chowdhury). The total time of $\approx 510$-hours was divided approximately equally between the seven DEEP2 subfields (21, 22, 31, 32, 33, 41 and 42); the total on-source time on each subfield was $\approx50-60$~hours. We used the GMRT Wideband Backend as the correlator, with a total bandwidth of 400~MHz covering the frequency range $530-930$~MHz, and divided into 8,192 spectral channels. 
	
 The data were analyzed in the Common Astronomy Software Applications package \citep[{\sc casa} Version 5; ][] {McMullin07} following standard procedures \citep[e.g.][]{Chowdhury20} to produce the final continuum images and spectral cubes for each of the seven DEEP2 subfields. The {\sc aoflagger} package \citep{Offringa12} was used to excise data affected by radio frequency interference (RFI). For each DEEP2 subfield, we chose to carry out an independent analysis of the data from the different GMRT observing cycles. Treating the data of different cycles independently prevents systematic errors (due to, e.g., low-level RFI, imperfect deconvolution, etc) in the data of one cycle from affecting the entire data set.
	
Our analysis yielded a total of 19 spectral cubes, two each for DEEP2 subfields 21 and 22 (which were each observed in only two GMRT cycles), and three each for DEEP2 subfields 31, 32, 33, 41, and 42 (which were each observed in all three GMRT cycles). The spectral cubes have a channel width of 48.8~kHz, corresponding to a velocity resolution of $18-25$~\kmps\ across our frequency coverage. The FWHMs of the synthesized beams of the cubes (i.e. the angular resolutions of the cubes) are $4.0''-7.5''$, corresponding to spatial resolutions of $29-63$~kpc for the redshift range $z=0.74-1.45$. 
	
The analysis also yielded 655~MHz continuum images of the seven DEEP2 subfields, with Root Mean Square (RMS) noise values of $\approx5-10~\mu$Jy~Beam$^{-1}$ in the central regions of the images, and synthesized beam FWHMs of $\approx 3''-4''$ \citep{Chowdhury22}.
	
	\subsection{HI 21 cm Emission Stacking}
		\label{sec:stackingHI}

 The upgraded GMRT observations covered the redshifted \hii\ line for 16,250 galaxies at $z=0.74-1.45$, for which the \hii\ lines are redshifted to $580-820$~MHz, and that lie within the full width at half maximum (FWHM) of the GMRT primary beam at the redshifted \hii\ line frequency of the galaxy. The FWHM of the GMRT primary beam was assumed to be $\approx43'$ at 610~MHz; the primary-beam FWHM scales $\propto 1/\nu$ with the observing frequency, $\nu$.

We restricted the \hii\ stacking to blue, star-forming galaxies, and hence excluded red galaxies and galaxies hosting active galactic nuclei (AGNs) from the sample. We excluded 2,222 red galaxies by only retaining objects with ${\rm C}\leq 0$, where ${\rm C=U-B}+0.032 \times ({\rm M_B}+21.63)-1.014$ \citep{Willmer06,Chowdhury20}. For AGNs, we excluded 882 DEEP2 objects that were detected at $\geq 4\sigma$ significance in our radio-continuum images, with 1.4~GHz luminosity L$_{1.4\ \textrm{GHz}} \geq 2\times10^{23}$~W~Hz$^{-1}$ \citep{Condon02}. 
We also excluded 487 galaxies with stellar mass $\Ms < 10^9 \ \Msun$, to ensure that our results can be directly compared with results for the xGASS survey at $z \approx 0$ \citep{Catinella18}. Finally, we carried out a suite of statistical tests on the \hii\ spectra of individual galaxies, to exclude any galaxies whose spectra are affected by non-Gaussian systematic effects \citep{Chowdhury20,Chowdhury22}. 

After the above exclusions, the main sample of the GMRT-CAT$z$1 survey contains 11,419 blue star-forming galaxies with accurate redshifts \citep[redshift quality, Q$\ge$3, in the DEEP2 DR4 catalogue; ][]{Newman13} at $z=0.74-1.45$, for which  the \hii\ line is redshifted to $\approx 580-820$~MHz (i.e. to frequencies where the Band-4 receivers have a high sensitivity), and that lie within the FWHM of the GMRT primary beam at the redshifted \hii\ frequency of the galaxy. The galaxies have stellar masses $\Ms \approx 10^9-10^{11}~\Msun$.
	
 For each galaxy, we made spectral subcubes, centred on the galaxy's position and redshift, and covering $\pm 500$~kpc around the galaxy position, and the velocity range $\pm 1500$~\kmps\ around the galaxy redshift. Each subcube was resampled onto a grid in the galaxy's rest frame; the resampled subcubes have spatial and velocity resolutions of 90~kpc and 30~\kmps, respectively. We also convolved the subcubes to coarser spatial resolutions ($\leq 200$~kpc) and find no evidence that the \hii\ emission is spatially resolved at the resolution of 90~kpc \citep{Chowdhury22}.

Our observations yielded multiple \hii\ subcubes, with uncorrelated statistical noise, for nearly every galaxy of our sample.  For each galaxy, there are $2-3$ independent \hii\ subcubes from the different GMRT cycles. In addition, some galaxies lie in the overlap regions of our GMRT pointings \citep{Chowdhury22}, yielding additional independent \hii\ subcubes. Overall, we obtain a total of 28,993 independent \hii\ subcubes for the 11,419 blue star-forming galaxies of our sample.

 For each of the 28,993 \hii\ subcubes, the flux-density values were first scaled to correct for the primary beam response at the galaxy's location in the GMRT primary beam. This was done using an azimuthally-symmetric polynomial fit to measurements of the average primary beam response of the GMRT Band-4 receivers.  Next, we converted each subcube from flux density (${\rm S}$) units to luminosity density (${\rm L}$) units at the redshift of the DEEP2 galaxy by using the relation ${\rm L}=4\pi \ 	{\rm S} \ {\rm D_L}^2/(1+z)$, where ${\rm D_L}$ is the luminosity distance of the galaxy.

 The stacking of the \hii\ emission signals was done by performing a weighted average of the spectral subcubes,  using the same weight for each spatial and velocity pixel, for different galaxy subsamples, to obtain an average \hii\ spectral cube for each set of stacked galaxies. The weights assigned to each subcube were specific to the goal of the stacking, and were used to control either the redshift or the  stellar-mass distributions of the galaxies in different subsamples. The weights did not take into account the RMS noise of the individual spectral cubes. The determination of the weights for the \hii\ subcubes for the specific cases of the stellar-mass subsamples and the redshift subsamples are discussed later. 

Finally, we fitted a second-order polynomial to each spatial pixel of the stacked \hii\ cube, and then subtracted this out to obtain a residual spectral cube. The second-order polynomial fit was performed after excluding the central $\pm250$~\kmps\ velocity range. We note that the exclusion of the velocity channels in the central $\pm250$~\kmps\ range from the fit results in a slight increase in the spectral RMS noise of these channels. 

The RMS noise on the stacked \hii\ spectral cube was determined via Monte Carlo simulations. In these simulations, we circularly shifted the central velocity of each galaxy by a random offset in the range $\pm1500$~\kmps, and then stacked the velocity-shifted \hii\ subcubes. The above procedure was repeated to obtain 10$^4$ realizations of the stacked \hii\ subcube. The RMS noise on every spatial and velocity pixel of the final stacked \hii\ cube was then estimated from these $10^4$ Monte Carlo realizations.

 Finally, the average \hi\ mass of a given subsample of DEEP2 galaxies was obtained from the velocity-integrated \hii\ line luminosity density ($\int {\rm L_{HI} \ dV}$, in units of ${\rm Jy~Mpc^2}$~\kmps), measured from the stacked \hii\ cube, via the expression $\MHI=[1.86 \times 10^4 \times \int {\rm L_{HI} \  dV}]~\Msun$.	The integral was carried out after smoothing the stacked cube to a velocity resolution of 90~\kmps, and was measured from a contiguous range of spectral channels with $\geq 1.5\sigma$ significance at this velocity resolution.

\subsection{Stacking the Rest-frame 1.4~GHz Continuum Emission}
\label{sec:stackingcont}

The average SFRs of the different galaxy subsamples were measured by stacking their  rest-frame 1.4~GHz radio luminosities \citep[e.g.][]{White07,Bera18,Chowdhury20},  using the correlation between the far-infrared and radio luminosities \citep[e.g.][]{Condon92,Yun01}. The radio continuum stacking was performed by extracting a subimage of each galaxy from the full continuum image of its DEEP2 subfield, and then smoothing and regridding each subimage to a spatial resolution of 40~kpc (at the galaxy redshift) and a uniform grid with 5.2~kpc pixels, extending  $\pm 260$~kpc around each galaxy. We note that the spatial resolution of 40~kpc is much larger than the typical size, $\approx 8$~kpc, of the star-forming regions of galaxies at these redshifts \citep[e.g.][]{Trujillo04}.  The flux density values at the observed frequency in every pixel of each subimage were scaled to the corresponding rest-frame 1.4~GHz flux densities at the galaxy's redshift, assuming a spectral index of $\alpha=-0.8$ \citep{Condon92}, with flux density $S_{\nu} \propto \nu^{\alpha}$. Next, we converted the rest-frame flux-density (${\rm S_{1.4\ GHz}}$) of every pixel of each subimage to luminosity-density  (${\rm L_{1.4\ GHz}}$) units using the relation ${\rm L_{1.4\  GHz}}=4\pi \ 	{\rm S_{1.4\ GHz}} \ {\rm D_L}^2/(1+z)$, where ${\rm D_L}$ is the luminosity distance of the galaxy. The stacked rest-frame 1.4~GHz luminosity was computed by taking a weighted median of the rest-frame 1.4~GHz luminosities of the corresponding spatial pixels of the different subimages, across the subsample of galaxies; the weights were chosen such that each galaxy has the same effective weight as in the \hii\ stack of the same subsample. The median rest-frame 1.4~GHz luminosity was then converted to a median SFR using the relation SFR~$(\Msun/\textrm{yr}) = (3.7 \pm 1.1) \times 10^{-22} \times {\rm L_{1.4\ GHz} (W/Hz)}$ \citep[][scaled to a Chabrier IMF]{Yun01}. Finally, for each stacked image, we corrected the rest-frame 1.4~GHz luminosity for a small zero-point offset ($\approx 0.4 \ \mu$Jy/Beam), at the level of $2-10\%$ of the measured value; the offset was estimated by applying the above stacking procedure to sub-images at locations offset by $100''$ from the DEEP2 galaxies of the sample.


We note that our radio-derived SFR estimates are affected by uncertainties in the flux density scale of our radio-continuum images. These systematic uncertainties are typically  $\lesssim 10\%$ for the GMRT, for our calibration procedure. Thus, all quoted errors on our SFR values have been estimated by adding, in quadrature, a $10\%$ systematic uncertainty to the $1\sigma$ statistical uncertainty.

\section{Results and Discussion}

\begin{figure*}
	
	\centering
	\includegraphics[width=0.85\linewidth]{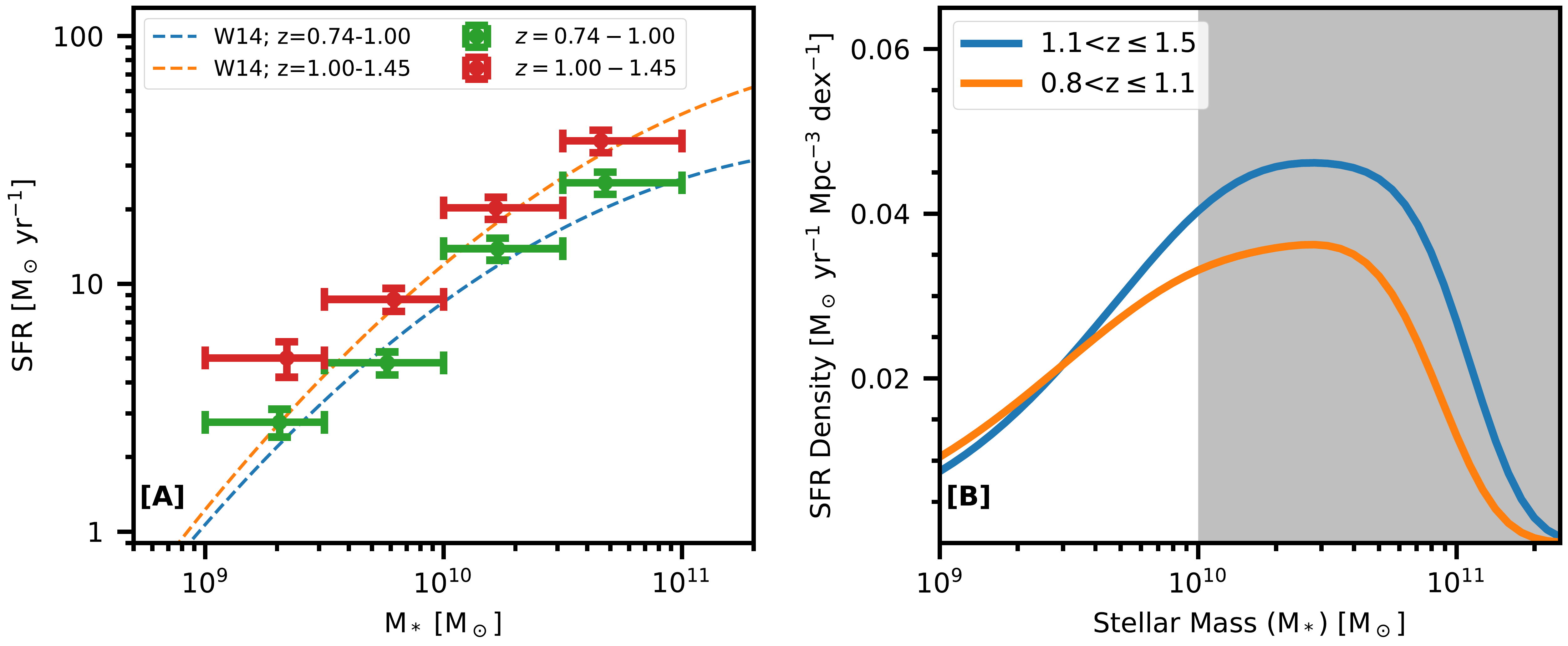}
	\caption{[A] The average SFR and the average stellar mass of our sample of 11,419 galaxies. The green and red points show, respectively, the average SFRs of the subsamples of galaxies at $z=0.74-1.00$ and $z=1.00-1.45$, in stellar mass bins of width 0.5~dex. The dotted lines show the star-forming main-sequence of  \cite{Whitaker14}, interpolated to the same redshift intervals. The average stellar mass and the average SFR of galaxies in each subsample are seen to be consistent with the star-forming main-sequence.  [B] The contribution of star-forming galaxies to the cosmic SFR density \citep[see also][]{Leslie20}. The figure shows the cosmic SFR density as a function of the stellar mass of galaxies in two redshift bins: (i)~during the epoch of peak cosmic SFR density, $z=1.1-1.5$ (blue), and (ii)~after the epoch of peak cosmic SFR density, $z=0.8-1.1$ (orange). Star-forming galaxies with $\Ms\geq 10^{10}\ \Msun$ (shown in the shaded region) dominate the SFR density at both epochs and contribute $\approx85\%$ to the decline in the cosmic SFR density at $z\lesssim1$.}
	\label{fig:sfrsd_Ms}
\end{figure*}
 Galaxies lying on the star-forming  main-sequence dominate the SFR density of the Universe out to $z\approx2.5$ \citep{Rodighiero11}. The decline in the SFR density of the Universe at $z\lesssim1$ is  caused by a decline in the zero-point of the star-forming main sequence \citep[e.g.][]{Whitaker14,Leslie20}.  We compared the stellar masses and SFRs of our sample of 11,419 galaxies with measurements of the star-forming main-sequence at these redshifts to determine if the sample galaxies are representative of main-sequence galaxies. This was done by dividing the sample of 11,419 galaxies into two redshift intervals, $z=0.74-1.00$ and $z=1.00-1.45$, and further dividing the galaxies in each redshift interval into multiple stellar-mass subsamples, each of width 0.5~dex. We then used the procedure of  Section~\ref{sec:stackingcont} to stack the rest-frame 1.4-GHz continuum emission of the galaxies in each subsample to determine their average SFRs. For the two redshift intervals, we used weights during the stacking process to ensure that the redshift distributions of the stellar-mass subsamples are identical.  The average stellar mass and the average SFR of each subsample, for both redshift intervals, are shown in Figure~\ref{fig:sfrsd_Ms}[A]. The figure also shows, for comparison, the star-forming main-sequence in the same redshift intervals, from \citet{Whitaker14}. Figure~\ref{fig:sfrsd_Ms}[A] shows that the average SFR and the average stellar mass of the galaxies in each subsample are consistent with the star-forming main sequence in both redshift intervals \citep[see also][]{Chowdhury22}. Our sample of 11,419 blue star-forming galaxies at $z=0.74-1.45$ is thus representative of main-sequence galaxies at these redshifts.

Detailed characterizations of the main sequence \citep{Whitaker14,Leslie20} and the stellar mass function of high-$z$ galaxies \citep{Davidzon17} have shown that the decline in the cosmic SFR density at $z \lesssim 1$ is predominantly due to the decline in the SFR density of the most massive galaxies, with stellar masses $\Ms \gtrsim 10^{10}\ \Msun$.  We combined the star-forming main sequence \citep{Whitaker14} and the stellar mass function of galaxies \citep{Davidzon17} to estimate the cosmic SFR density of galaxies as a function of their stellar mass. Figure~\ref{fig:sfrsd_Ms}[B] shows the cosmic SFR density of galaxies as a function of their stellar mass in two redshift intervals, with mean redshifts of $\langle z \rangle = 1.3$ and $\langle z \rangle = 0.95$, respectively. Star-forming galaxies with $\Ms \gtrsim 10^{10} \ \Msun$, shown as the shaded region in Figure~\ref{fig:sfrsd_Ms}[B], contribute $\approx 85\%$ to the decline in the total SFR density from $z\approx1.3$ to $z\approx0.95$.  
Maintaining the \hi\ reservoir in such galaxies would require accretion to take place at a high rate. If the lack of gas accretion in galaxies causes the decline of the cosmic SFR density at $z \lesssim 1$, one would expect that the most massive galaxies, which dominate the SFR density at these redshifts, would consume the bulk of their \hi\ reservoir on a very short timescale, comparable to the $\approx 1$-billion year interval between $z \approx 1.3$ and $z \approx 1.0$.

\subsection{Very Short \hi\ Depletion Timescale in Massive Galaxies}

\begin{figure*}
    \centering
    \includegraphics[width=0.7\linewidth]{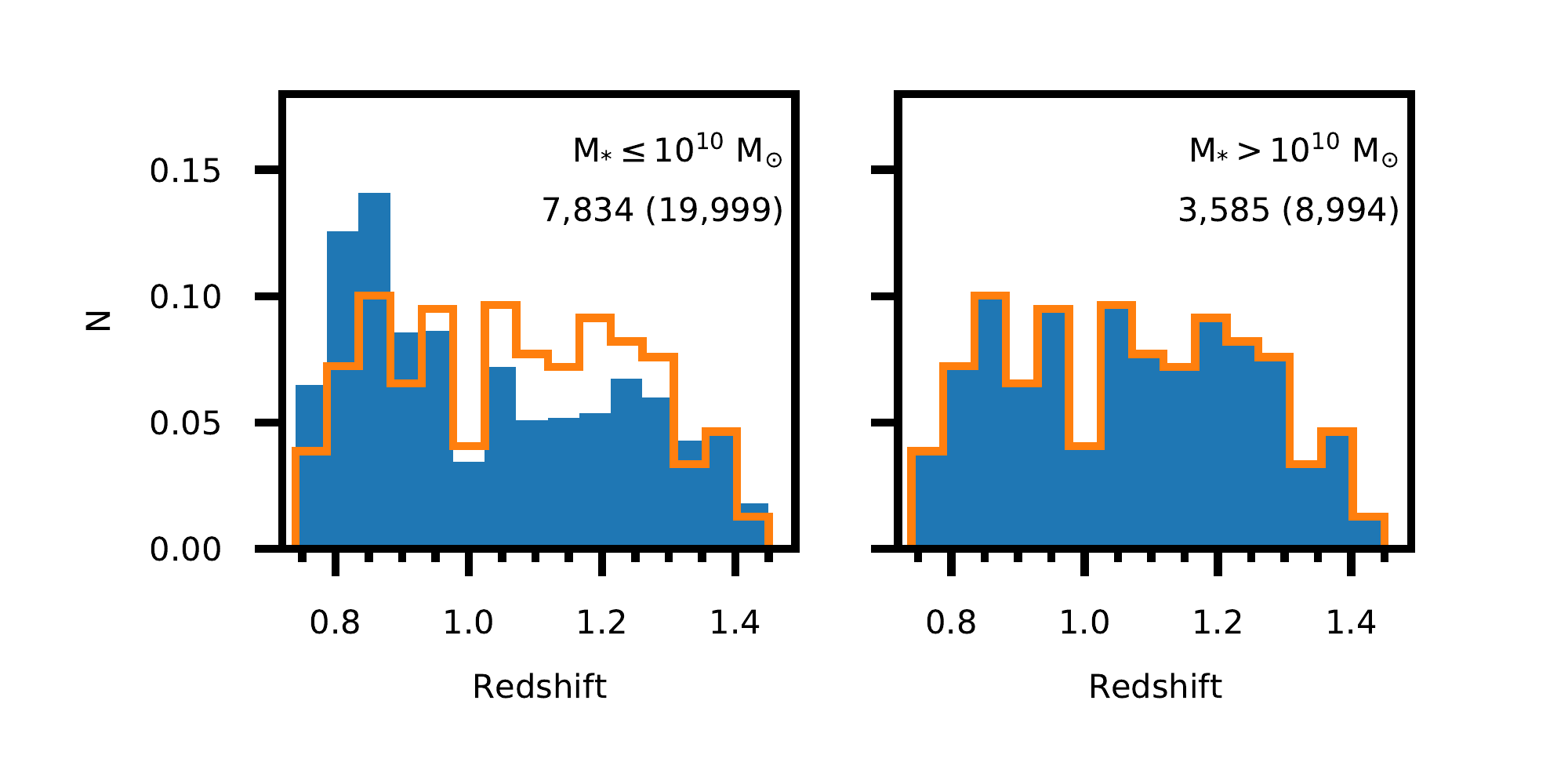}
     \vspace{-20pt}

        \caption{
    The redshift distributions of galaxies in the two stellar-mass subsamples. The blue histograms show the number of \hii\ subcubes as a function of redshift in each of the two stellar-mass subsamples; each histogram has been normalized by the total number of subcubes in its subsample. The orange histogram in each panel shows the redshift distribution of galaxies in the high stellar-mass subsample. The \hii\ subcubes of the lower stellar-mass subsample were assigned weights such that their effective redshift distribution is the one shown in orange, i.e. is identical to that of the higher stellar-mass subsample.  The number of galaxies in each subsample is shown above each histogram, with the number of \hii\ subcubes in parentheses.} 
    \label{fig:zdist}
\end{figure*}

We hence examined the dependence of the average \hi\ mass and the \hi\ depletion timescale on the average stellar mass, by dividing the full sample of 11,419 galaxies into two subsamples,  with [A]~$\Ms \leq 10^{10}\ \Msun$ (7,834 galaxies and 19,999 independent \hii\ subcubes) and [B]~$\Ms> 10^{10}\ \Msun$ (3,585 galaxies and 8,994 independent subcubes). Figure~\ref{fig:zdist} shows the redshift distributions of the \hii\ subcubes in each of the two stellar-mass subsamples. It is clear that the lower stellar-mass subsample contains more galaxies at lower redshifts than the higher stellar-mass subsample; this is because the DEEP2 survey is a magnitude-limited survey, and is hence biased toward detecting fainter (i.e., typically less massive) galaxies at lower redshifts. We account for this difference in the redshift distributions of the two stellar-mass subsamples by assigning weights to each galaxy to ensure that the effective redshift distribution of the two stellar-mass subsamples is identical. Specifically, we assigned weights to the \hii\ subcubes in the low stellar-mass subsample to make their effective redshift distribution identical to that of the \hii\ subcubes in the high stellar-mass subsample.  These weights were then used in stacking both the \hii\ emission and the rest-frame 1.4~GHz radio continuum of the galaxies of each subsample, following the procedures in Sections~\ref{sec:stackingHI} and \ref{sec:stackingcont}, to determine the average \hi\ mass and the average SFR of galaxies in the two subsamples. These weights were further used in the determination of all average quantities for the two stellar-mass subsamples. In passing, we note that the effective redshift distribution of the two subsamples has a mean of $\langle z \rangle \approx 1.1$.

\begin{figure*}
    \centering
    \includegraphics[width=\linewidth]{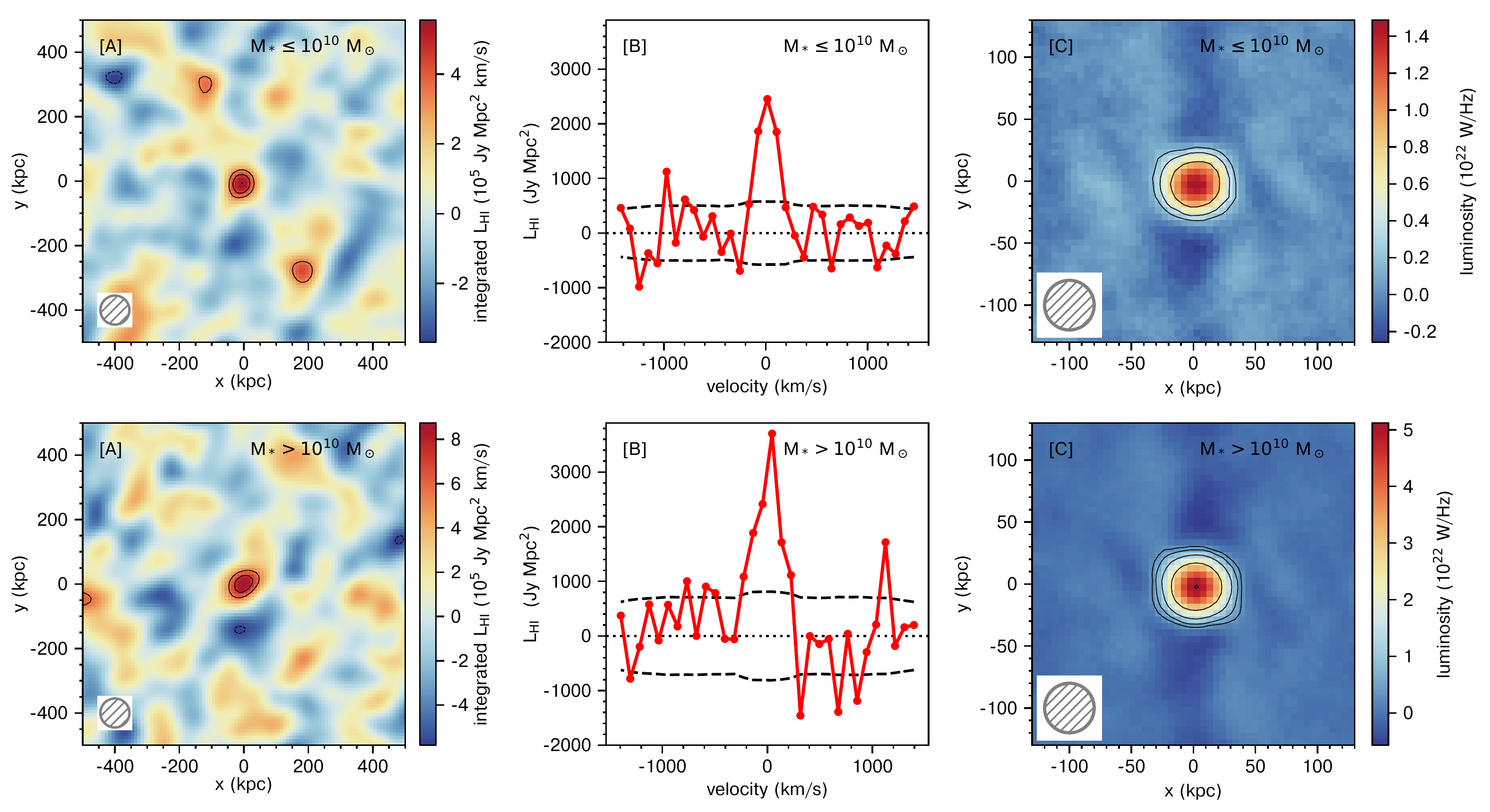}

    \caption{ The stacked \hii\ emission and the stacked rest-frame 1.4 GHz continuum emission from star-forming galaxies with different stellar masses at $z\approx1$. [A]~The left panels show, for the two stellar-mass subsamples, the average \hii\ emission images. The circle on the bottom left of each panel shows the 90-kpc spatial resolution of the images. The contour levels are at $-3.0\sigma$ (dashed), $+3.0\sigma$, $+4.0\sigma$, and $+5.0\sigma$ significance. [B]~The middle panels show the average \hii\ emission spectra for the two subsamples; the channel width is 90~\kmps, and the dashed curves show the $\pm1\sigma$ error on each spectrum. The average \hii\ emission signals from both subsamples are clearly detected in both the images and the spectra. [C]~The right panels show the stacked rest-frame 1.4~GHz luminosity images of the galaxies in the two subsamples. The contour levels are at $5\sigma,\ 10 \sigma,\ 20 \sigma,\ 40 \sigma,\ {\rm and} \ 80\sigma$ statistical significance. The circle at the bottom left of each panel shows the 40~kpc resolution of the images. The average rest-frame 1.4 GHz continuum emission is detected at high ($> 37\sigma$) significance in both images. }
    \label{fig:msstacks}
\end{figure*}
\begin{deluxetable*}{lcccccc}
	\tablenum{1}
	\tabletypesize{}

	\tablecaption{Average properties of galaxies in the stellar mass and the redshift subsamples. }   
	\tablehead{\colhead{\,}  & \colhead{\hspace{2.5cm}\,} & \multicolumn2c{Stellar-mass Subsamples} & \colhead{\hspace{2.5cm}\,} & \multicolumn2c{Redshift Subsamples}  \\
		\cline{3-4} \cline{6-7}
		\colhead{\,}  &\colhead{\,}  & \colhead{Low} & \colhead{High} & \colhead{\,}  & \colhead{Low} & \colhead{High}
	}
	
	\startdata
	Stellar Mass Range ($\times 10^{9}\ \Msun$) & & $1.0-10$  & $10-240$  & & $1.0-240$  & $1.0-240$ \\
	Redshift Range & & $0.74-1.45$  & $0.74-1.45$  & & $0.74-1.25$ & $1.25-1.45$ \\
	Number of Galaxies & & $7,834$ & $3,585$  & & $9,284$ & $2,135$\\
	Number of \hii\ subcubes & & $19,999$ & $8,994$ & & 23,630 & 5,363\\
	Average Redshift, $\langle z \rangle$  & & $1.1$  & $1.1$ & & 0.98 & 1.33 \\
	Average Stellar Mass ($\times 10^{9}\ \Msun$) & & $4.9$  & $21.4$ & & 10.3 & 10.3  \\
	Average H{\sc i} Mass  ($\times 10^{9}\ \Msun$) & & $10.4\pm2.1$ &  $16.3\pm3.5$  & & $10.6\pm1.9$ & $33.6\pm6.4$ \\
	Average SFR ($\Msun\textrm{yr}^{-1}$)& & $5.5\pm0.6 $ & $18.9\pm1.9$& & $7.5\pm0.8 $ & $15.8\pm1.7$\\
	Average H{\sc i} depletion timescale (Gyr) & & $1.93\pm0.44$  & $0.86\pm0.20$  & & $1.41 \pm 0.31$ & $2.01 \pm 0.37$\\[0.2cm]
	\enddata
	\tablecomments{For each of the two stellar mass subsamples (columns~2 and 3) and the two redshift subsamples (columns~4 and 5), the rows are (1)~the range of stellar masses, (2)~the number of galaxies in each subsample, (3)~the number of \hii\ subcubes in each subsample, (4)~the average redshift of the galaxies in each subsample, after applying the weights, (5) the average stellar mass of the galaxies in each subsample, after applying the weights, (6)~the average H{\sc i} mass of the galaxies in each subsample, (7)~the average SFR of the galaxies in each subsample, and (8) the average H{\sc i} depletion timescale, $\langle {\rm t}_{\rm dep,H\textsc{i}} \rangle \equiv \langle \MHI \rangle/\langle {\rm SFR} \rangle$, of the galaxies in each subsample. See the main text for a description of the weights that were applied to different subsamples.  We note that the distribution of the galaxies within the GMRT primary beam, as a function of the distance from the pointing centre, is very similar for all four subsamples.}
	\label{tab:insuffacc:msstacks}
\end{deluxetable*}
Figure~\ref{fig:msstacks} shows the stacked \hii\ emission image, the stacked \hii\ spectrum, and the stacked rest-frame 1.4~GHz continuum image of the galaxies in each of the two stellar-mass subsamples. We obtain clear detections of the average \hii\ emission signal from both subsamples, each at $>4.6\sigma$ statistical significance, as well as detections of the rest-frame 1.4~GHz luminosity, each at $> 37\sigma$ significance. This allows us to  determine the average \hi\ mass and the average SFR of the galaxies of each stellar-mass subsample.

The \hi\ depletion timescale, $\textrm{t}_{\textrm{dep,H}\textsc{i}}=\textrm{M}_{\textrm{H}\textsc{i}}$/SFR, quantifies the timescale on which a galaxy would convert its entire \hi\ reservoir into stars, if continuing to form stars at its current SFR. Star-formation would be quenched in a galaxy on timescales longer than  $\approx \textrm{t}_{\textrm{dep,H}\textsc{i}}$, if the \hi\ reservoir is not replenished by accretion of gas from the circumgalactic medium or by minor mergers. We combined our estimates of the average \hi\ mass and the average SFR in the two stellar-mass subsamples to estimate $\langle \textrm{t}_{\textrm{dep,H}\textsc{i}}\rangle \equiv \langle \textrm{M}_{\textrm{H}\textsc{i}} \rangle / \langle {\rm SFR} \rangle$ for each subsample. The inferred average stellar mass, average \hi\ mass, average SFR, and average \hi\ depletion timescale of the two stellar-mass subsamples are listed in the last four rows of  Table~\ref{tab:insuffacc:msstacks}.

\begin{figure}
	\centering
	\includegraphics[width=0.6\linewidth]{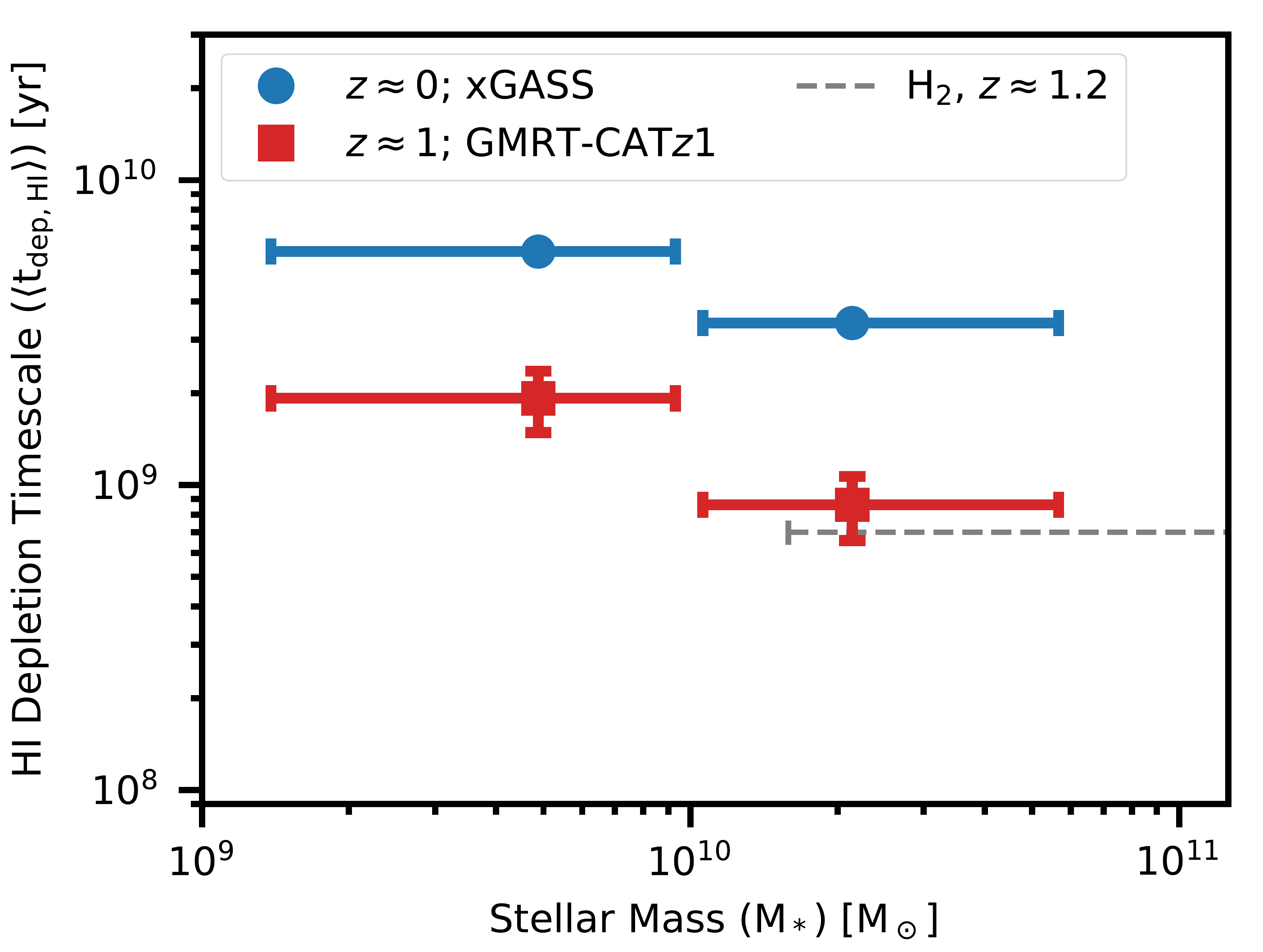}
	
	\caption{The average \hi\ depletion timescale, $\langle {\rm t}_{\rm dep,H\textsc{i}} \rangle \equiv \langle \MHI \rangle/\langle {\rm SFR} \rangle$, of star-forming galaxies at $z\approx1.1$, plotted against average stellar mass. The red circles show our measurements of the average \hi\ depletion timescale for the two stellar mass subsamples. The error bars along the y-axis indicate the $1\sigma$ statistical uncertainties on the measurements while the error bars along the x-axis show the 90\% range of stellar masses in each subsample. The blue points show the average \hi\ depletion timescale, $\langle {\rm t}_{\rm dep,H\textsc{i}} \rangle \equiv \langle \MHI \rangle/\langle {\rm SFR} \rangle$, for two subsamples of xGASS galaxies at $z \approx 0$ \citep{Catinella18}, each with the same stellar-mass distribution as the corresponding DEEP2 subsample. The dashed grey line shows the $\htwo$ depletion timescale of star-forming galaxies at $z\approx1.2$ \citep{Tacconi13}. We obtain a very short average \hi\ depletion timescale, $\approx 0.86$~Gyr, for star-forming galaxies with $\Ms > 10^{10} \ \Msun$ at $z \approx 1.1$, comparable to the $\htwo$ depletion timescale \citep{Tacconi13} in similar galaxies at $z \approx 1.2$.}
	\label{fig:tdepMs}
\end{figure}

Figure~\ref{fig:tdepMs} plots the average \hi\ depletion timescale of the two stellar-mass subsamples against their average stellar mass. For comparison, we also plot $\langle \textrm{t}_{\textrm{dep,H}\textsc{i}} \rangle$ against average stellar mass for a reference sample of $z \approx 0$ galaxies, the stellar mass-selected xGASS sample \citep{Catinella18}. For the xGASS galaxies, we have restricted to blue galaxies, with ${\rm NUV-r} < 4$, and used weights to obtain an effective stellar-mass distribution matched to that of the DEEP2 subsamples. We find that the average \hi\ depletion timescale of DEEP2 galaxies with $\Ms \le 10^{10} \ \Msun$ is $\approx2$~Gyr, a factor of $\approx 3$ smaller than that of similar galaxies in the local Universe. However, remarkably, we find a far shorter average \hi\ depletion timescale,  $\langle \textrm{t}_{\textrm{dep,H}\textsc{i}}\rangle = 0.86\pm0.20$~Gyr, in the most massive DEEP2 galaxies, with $\Ms > 10^{10} \ \Msun$, at $z \approx 1.1$.  This is comparable to the $\approx 1.0$~Gyr period between $z = 1.3$ and $z = 1$ (corresponding to lookback times of $\approx 9$~Gyr and 8~Gyr, respectively), and to the molecular gas depletion timescale, $\textrm{t}_{\rm dep,H_2} \approx 0.7$~Gyr, for similar galaxies at these redshifts \citep{Tacconi13}. Star-forming galaxies with $\Ms > 10^{10} \ \Msun$ would thus be able to sustain their SFR for only about a billion years unless their \hi\ is replenished on this timescale via accretion from the circumgalactic medium or mergers. The very short \hi\ depletion timescale in the most massive galaxies, that dominate the decline in the cosmic SFR density at $z \lesssim 1$, is consistent with the hypothesis that most of the \hi\ in such galaxies was consumed by $z \approx 1$, and that the lack of accretion of \hi\ onto such galaxies led to the quenching of their star-formation and hence, to the decline in the cosmic SFR density at $z \lesssim 1$.
\subsection{Rapid Redshift Evolution in the \hi\ mass of Star-forming Galaxies over $z \approx 0.74-1.45$}

The redshift coverage of the GMRT CAT$z$1 survey covers both the peak of cosmic SFR density at $z \gtrsim 1$ and the decline of the SFR density at $z\lesssim1$. This allows us to directly test the hypothesis that insufficient gas accretion caused the decline in the cosmic SFR density, by measuring the \hi\ content of star-forming galaxies of the same stellar mass in two redshift bins, with average redshifts $\langle z\rangle \approx 1.3$ and $\langle z \rangle \approx 1$. If the \hi\ reservoir of star-forming galaxies at $z\approx 1.3$ is indeed not replenished by accretion, we expect to observe a significant decline in the average \hi\ mass of such galaxies, from $\langle z \rangle = 1.3$ to $\langle z \rangle = 1$.

\begin{figure*}
    \centering
    \includegraphics[width=0.7\linewidth]{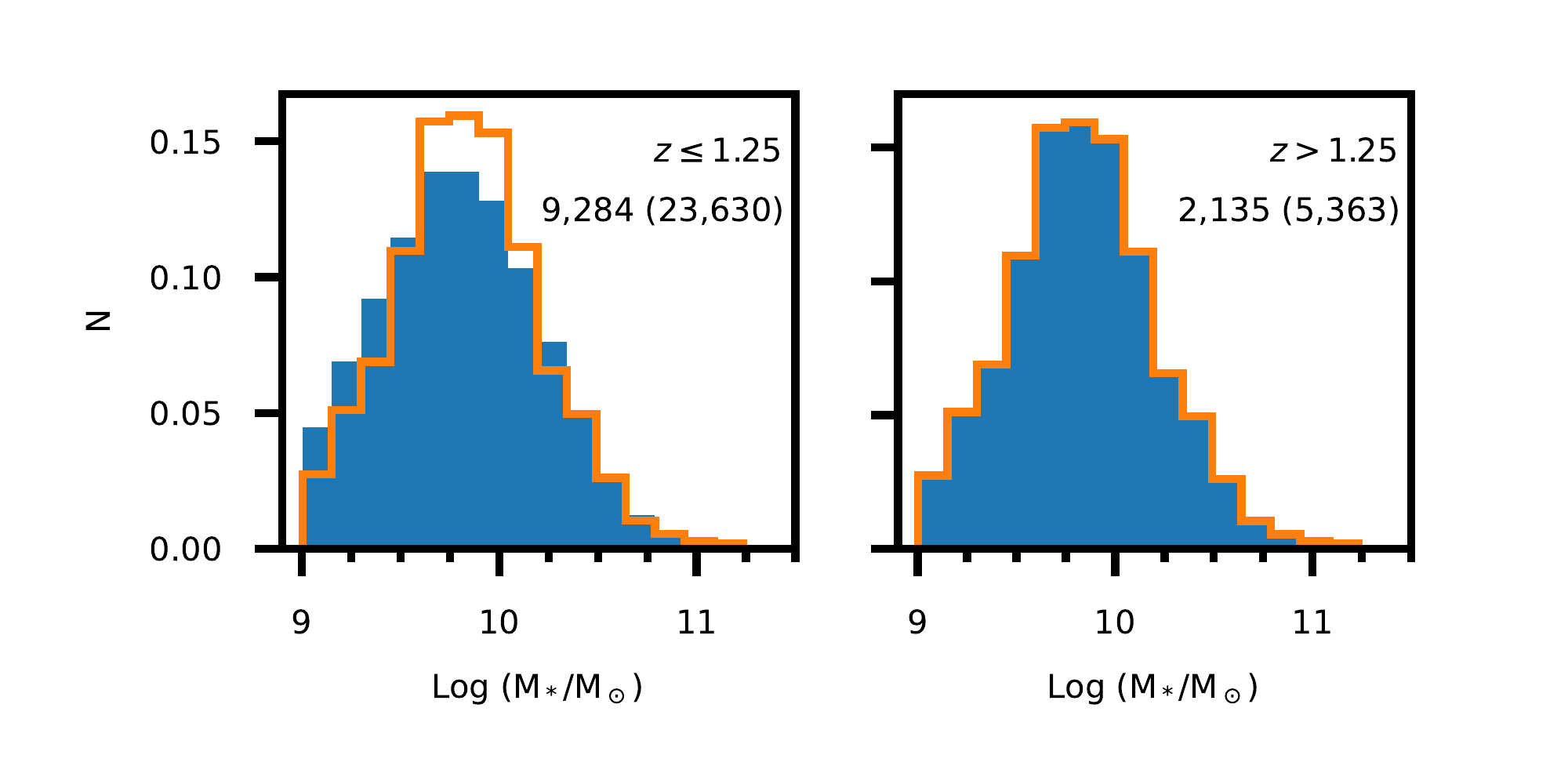}
    \vspace{-20pt}
  
    \caption{The stellar-mass distributions of galaxies in the two redshift subsamples. The blue histogram shows the number of \hii\ subcubes, in bins of stellar mass, in the two redshift subsamples; each histogram has been normalized by the total number of subcubes in its subsample. The orange histograms show the stellar-mass distribution of galaxies at $z=1.25-1.45$. The galaxies in the lower-redshift subsample were assigned weights such that their effective stellar-mass distribution is the one shown in orange, i.e. is identical to the stellar-mass distribution of galaxies in the higher-redshift subsample. The number of galaxies in each subsample is shown above the histogram, with the number of \hii\ subcubes shown in parentheses.}
    \label{fig:insuffacc:msDist}
\end{figure*}
To test this hypothesis, we divided our sample of 11,419 star-forming galaxies into two subsamples based on the galaxy redshift. The high-redshift sample includes 2,135 galaxies (5,363 independent \hii\ subcubes) at $z = 1.25-1.45$ (i.e. during the epoch of the peak of cosmic SFR density, with an average redshift $\langle z \rangle \approx 1.3$), while the low-redshift sample contains 9,284 galaxies (23,630 independent subcubes) at $z = 0.74 - 1.25$ (i.e. covers the decline of the cosmic SFR density, with $\langle z \rangle \approx 1.0$). The intrinsic stellar-mass distributions of the two subsamples, shown in Figure~\ref{fig:insuffacc:msDist}, are different. We corrected for this effect by assigning weights to each galaxy in the lower-redshift subsample to ensure that the stellar-mass distributions of the two subsamples are identical; the weighted-mean stellar mass of each subsample is $\approx 10^{10}\ \Msun$. 

\begin{figure*}
    \centering

    \includegraphics[width=\linewidth]{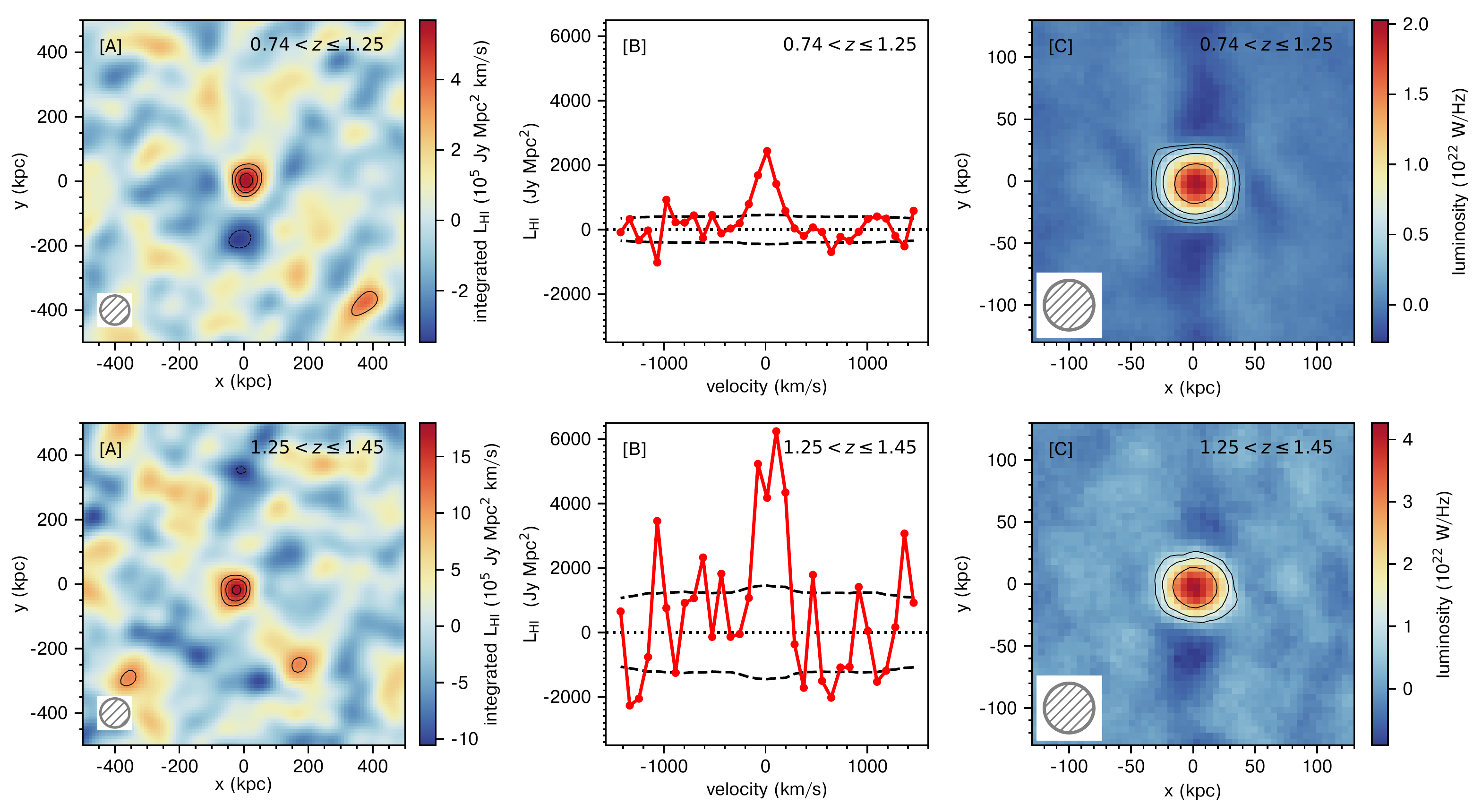}
    \scriptsize
    \caption{The average \hii\ emission and the average rest-frame 1.4 GHz continuum emission from star-forming galaxies in the two redshift subsamples. [A]~The left panels show, for the two stellar-mass subsamples, the stacked \hii\ emission images. The circle at the bottom left of each panel shows the 90-kpc spatial resolution of the images. The contour levels are at $-3.0\sigma$ (dashed), $+3.0\sigma$, $+4.0\sigma$, and $+5.0\sigma$ significance. [B]~The middle panels show the average \hii\ emission spectra; the channel width is 90~\kmps, while the dashed curves show the $\pm1\sigma$ error on each spectrum. The stacked \hii\ emission signals from both subsamples are clearly detected in both the images and the spectra. [C]~The right panels show the stacked rest-frame 1.4~GHz luminosity images of the galaxies in the two subsamples.  The contour levels are at $5\sigma,\ 10 \sigma,\ 20 \sigma, {\rm and} \ 40\sigma$ statistical significance. The circle at the bottom left of each panel shows the 40~kpc resolution of the images. The average rest-frame 1.4 GHz continuum emission is clearly detected at high ($> 32\sigma$) significance in both images.  }
    \label{fig:zstacks}
\end{figure*}

Using the above weights, we stacked the \hii\ subcubes and the rest-frame 1.4~GHz continuum subimages of the galaxies in the two redshift subsamples, following the procedures of Section~\ref{sec:stackingHI} and \ref{sec:stackingcont}. This yielded clear detections of the average \hii\ emission signal (with $>5.2\sigma$ statistical significance) and the average rest-frame 1.4~GHz luminosity (with $> 32\sigma$ statistical significance),
from both subsamples.  Figure~\ref{fig:zstacks} shows the stacked \hii\ emission image, the stacked \hii\ spectrum, and the stacked rest-frame 1.4~GHz radio-continuum image of the  galaxies in the two redshift subsamples. We find that star-forming galaxies with $\langle \Ms \rangle \approx 10^{10} \ \Msun$ at $\langle z \rangle \approx 1.0$  have an average \hi\ mass of $\MHI=(10.6 \pm 1.9) \times10^{9} \  \Msun$, while those at $\langle z \rangle \approx 1.3$ have an average \hi\ mass of $\MHI=(33.6 \pm 6.4) \times 10^{9} \ \Msun$. We thus find evidence, at $3.4\sigma$ statistical significance, that the average \hi\ mass of star-forming galaxies with $\langle \Ms \rangle \approx 10^{10}~\Msun$ at $\langle z \rangle \approx 1.3$ is higher, by a factor of $\approx 3.2$, than that of galaxies with an identical stellar-mass distribution at $\langle z \rangle \approx 1.0$.
Finally, stacking the 1.4~GHz luminosities of the two subsamples yields average SFRs of $(15.8 \pm 1.6) \ \Msun$~yr$^{-1}$ at $\langle z \rangle \approx 1.3$, but only $(7.5 \pm 0.8) \ \Msun$~yr$^{-1}$ at $\langle z \rangle \approx 1.0$.  We thus find direct evidence that the \hi\ reservoir of star-forming galaxies is indeed rapidly depleted between $z \approx 1.3$ and $z \approx 1.0$, i.e. on a timescale of $\approx 1$~billion years. The decline in \hi\ mass is by a factor of $(3.2 \pm 0.8)$, consistent with the decline in the average SFR, by a factor of $(2.1 \pm 0.3)$, of the same galaxies over the same redshift interval. 

\begin{figure}
    \centering
    \includegraphics[width=0.6\linewidth]{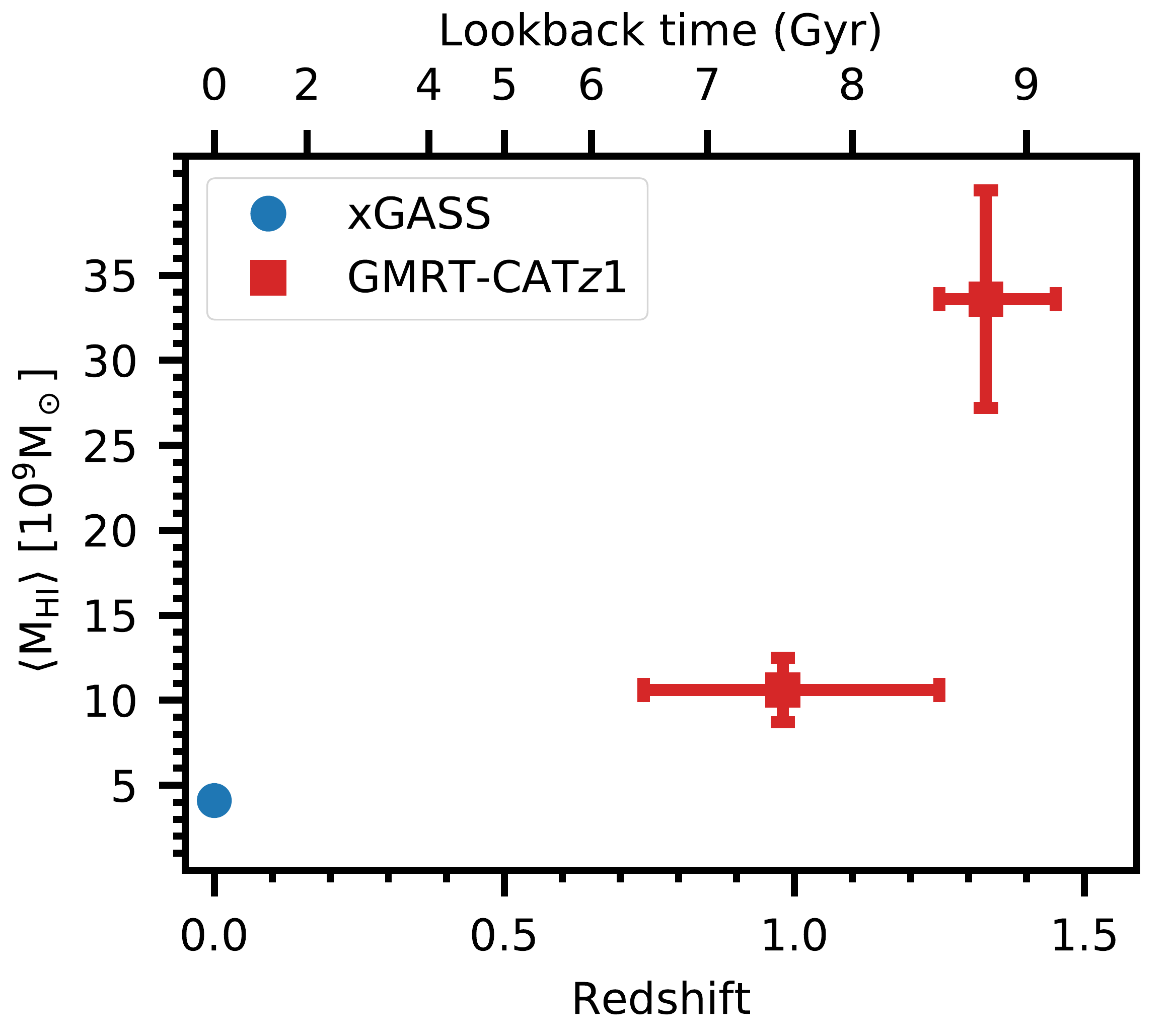}
    \caption{The redshift evolution of the average \hi\ mass of star-forming galaxies over $z = 0.74-1.45$. The red squares show our measurements of the average \hi\ mass of star-forming galaxies in the two redshift subsamples at $\langle z \rangle \approx 1.3$ and $\langle z \rangle \approx 1.0$, with identical stellar mass distributions, obtained by separately stacking the \hii\ signals of galaxies in each subsample. The average \hi\ mass of blue xGASS galaxies at $z \approx 0$ \citep{Catinella18}, with an identical effective stellar mass distribution, is marked with a blue circle. The average \hi\ mass is seen to decline steeply from $z \approx 1.3$ to $z \approx 1.0$, and to then decline slowly down to $z \approx 0$, for galaxies of the same average stellar mass,  $\langle \Ms \rangle \approx 10^{10}\ \Msun$.}
    \label{fig:MHI_z}
\end{figure}

Figure~\ref{fig:MHI_z} plots the average \hi\ mass of the DEEP2 galaxies in the two redshift subsamples against redshift, along with the average \hi\ mass of blue xGASS galaxies at $z \approx 0$, again with weights chosen to yield the same stellar-mass distribution as that of the high-$z$ galaxies. The average \hi\ mass\footnote{The error on the average \hi\ mass of the xGASS galaxies was computed using bootstrap resampling with replacement.} of blue xGASS galaxies with $\langle \Ms \rangle \approx 10^{10} \ \Msun$ is $(4.1 \pm 0.2) \times 10^9  \ \Msun$.  This is a factor of $\approx 2.6$ lower than the average \hi\ mass of our lower-redshift subsample, $\MHI = (10.6 \pm 1.9) \times 10^9 \ \Msun$ at $\langle z \rangle \approx 1.0$. We thus also find evidence (at $\approx 3.4\sigma$ significance) for evolution in the average \hi\ mass of star-forming galaxies with the same stellar-mass distribution between $z \approx 1$ and $z \approx 0$. Overall, Figure~\ref{fig:MHI_z} shows that the average \hi\ mass in galaxies with $\langle \Ms \rangle \approx 10^{10} \ \Msun$ rapidly declines, by a factor of $\approx 3.2$, over a period of $\approx 1$~billion years from $z \approx 1.3$ and $z \approx 1.0$, and then gradually declines, by a factor of $\approx2.6$, over the next $\approx 8$~billion years from $z\approx1$ to $z\approx0$.


Finally, we investigated the possibility that our measurement of the decline in the average \hi\ mass of star-forming galaxies from $z\approx1.3$ to $z\approx1.0$ might be caused by incompleteness issues in the DEEP2 sample. First, we note that the galaxies in both redshift subsamples are representative of the main-sequence population at their redshifts (see Figure~\ref{fig:sfrsd_Ms}[A]). We tested for incompleteness issues by restricting our analysis of the redshift evolution of the \hi\ mass to only the most massive galaxies of the sample, with stellar masses greater than the median stellar mass, $6.3 \times 10^9 \ \Msun$; such galaxies are much less likely to be affected by incompleteness. We divided this massive-galaxy sample of 5,751 galaxies with $\Ms>6.3\times10^9~\Msun$ into two redshift subsamples at $z=0.74-1.25$ (4,589 galaxies) and $z=1.25-1.45$ (1,162 galaxies) and separately stacked the \hii\ subcubes of the galaxies in each redshift subsample, following the procedures of Section~\ref{sec:stackingHI}, 
again ensuring that the stellar-mass distribution of the two subsamples are identical. We find that the average \hi\ masses
of star-forming galaxies with 
$\langle\Ms\rangle\approx 1.6 \times 10^{10} 
\ \Msun$ at $\langle z\rangle\approx1.0$ and $\langle z \rangle\approx1.3$ are $\MHI=(16.0 \pm 3.2) \times10^{9} \  \Msun$ and $\MHI=(48 \pm 10) \times10^{9} \  \Msun$, respectively. 
Thus, even for the massive-galaxy subsample, we continue to find evidence, at $\approx3.0\sigma$ significance, for a decline in the average \hi\ mass between $z \approx 1.3$ and $z \approx 1.0$. Overall, we do not find any evidence that the measurement of the redshift evolution of the average \hi\ mass of main-sequence galaxies presented in this \emph{Letter} might be affected by selection biases.

\section{Summary}
We report first results from the GMRT-CAT$z$1 survey, a $510$-hour \hii\ emission survey of galaxies at $z=0.74-1.45$ in the DEEP2 fields. We use stacking of the \hii\ emission and the rest-frame 1.4~GHz radio-continuum of subsamples of DEEP2 galaxies to determine the dependence of the average \hi\ mass and the average \hi\ depletion time of main-sequence galaxies at $z=0.74-1.45$ on their average stellar mass and redshift. We find that \hi\ plays a fundamental role in the decline of the cosmic SFR density at $z\lesssim1$. The very short average \hi\ depletion timescale of $\approx 0.86$~billion years in the most massive galaxies, which dominate the cosmic SFR density at $z > 1$, implies that rapid gas accretion is needed to maintain their \hi\ reservoirs and sustain their high SFRs. And the sharp decline in the average \hi\ mass of galaxies with $\langle \Ms \rangle \approx 10^{10} \ \Msun$, by a factor of $\approx 3.2$ from $z \approx 1.3$ to $z \approx 1$, indicates that such efficient gas accretion does not take place. We conclude that the lack of accretion of \hi\ onto the most massive star-forming galaxies towards the end of the epoch of peak cosmic SFR density causes the decline in the cosmic SFR density at $z \lesssim 1$, eight billion years ago.

	\begin{acknowledgments}
	We thank the staff of the GMRT who have made these observations possible. The GMRT is run by the National Centre for Radio Astrophysics of the Tata Institute of Fundamental Research. We thank an anonymous referee for suggestions which improved the clarity of the paper. NK acknowledges support from the Department of Science and Technology via a Swarnajayanti Fellowship (DST/SJF/PSA-01/2012-13). AC, NK, $\&$ JNC also acknowledge the Department of Atomic Energy for funding support, under project 12-R\&D-TFR-5.02-0700. 
	\end{acknowledgments}
    \software{CASA \citep{McMullin07},   
      calR \citep{calR},  AOFLAGGER \citep{Offringa12}, 
      astropy \citep{astropy:2013,astropy:2018}}
      
    \bibliography{bibliography.bib}

\bibliographystyle{aasjournal}

\end{document}